# Pressure dependent relaxation in the photo-excited Mott insulator ET-F$_2$TCNQ: influence of hopping and correlations on quasiparticle recombination rates

M. Mitrano$^{(1)}$, G. Cotugno$^{(1,2)}$, S.R. Clark$^{(3,2)}$, R. Singla$^{(1)}$, S. Kaiser$^{(1)}$, J. Stähler$^{(4)}$, R. Beyer$^{(5)}$, M. Dressel$^{(5)}$, L. Baldassarre$^{(6)}$, D. Nicoletti$^{(1)}$, A. Perucchi$^{(7)}$, T. Hasegawa$^{(8)}$, H. Okamoto$^{(9)}$, D. Jaksch$^{(2,3)}$, A. Cavalleri$^{(1,2)}$

$^{(1)}$Max Planck Institute for the Structure and Dynamics of Matter, Luruper Chaussee 149, Hamburg, Germany

$^{(2)}$ Department of Physics, Oxford University, Clarendon Laboratory, Parks Road, Oxford, United Kingdom

$^{(3)}$Centre for Quantum Technologies, National University of Singapore, Singapore

$^{(4)}$Fritz Haber Institute of the Max Planck Society, Faradayweg 4-6, 14195 Berlin, Germany

$^{(5)}$1. Physikalisches Insitut, Universität Stuttgart, Pfaffenwaldring 57, 70550 Stuttgart, Germany

$^{(6)}$ Center for Life NanoScience@Sapienza, Istituto Italiano di Tecnologia, V.le Regina Elena 291, Rome, Italy

$^{(7)}$ INSTM UdR Trieste-ST and Elettra - Sincrotrone Trieste S.C.p.A., S.S. 14 km 163.5 in Area Science Park, 34012 Basovizza, Trieste Italy

$^{(8)}$National Institute of advanced Industrial Science and Technology, Tsukuba, Japan

$^{(9)}$Department of Advanced material Science, University of Tokyo, Chiba 277-8561, Japan

We measure the ultrafast recombination of photo-excited quasiparticles (holon-doublon pairs) in the one dimensional Mott insulator ET-F$_2$TCNQ as a function of external pressure, which is used to tune the electronic structure. At each pressure value, we first fit the static optical properties and extract the electronic bandwidth $t$ and the intersite correlation energy $V$. We then measure the recombination times as a function of pressure, and we correlate them with the corresponding microscopic parameters. We find that the recombination times scale differently than for metals and semiconductors. A fit to our data based on the time-dependent extended Hubbard Hamiltonian suggests that the competition between local recombination and delocalization of the Mott-Hubbard exciton dictates the efficiency of the recombination.

The recombination of hot carriers in solids is a fundamental process of interest to nonlinear optics and to device applications, as well as a spectroscopic tool that exposes the physics of interacting microscopic degrees of freedom. "Hot electron" spectroscopy has been applied extensively to metals and semiconductors, for which well-established models have been developed.

For direct gap semiconductors recombination occurs at a rate that depends on the joint density of states between valence and conduction bands $\propto \frac{\partial E_v}{\partial k}\frac{\partial E_c}{\partial k}$, and is thus expected to *slow down* with the square of the bandwidth $\tau \propto t^2$. On the other hand, in the case of metals, the dynamics are well captured by the two-temperature model [1,2], which considers the energy stored in the optically excited non-equilibrium electron distribution as flowing into the lattice at a rate determined by the electron-phonon coupling strength and by the electronic and lattice heat capacities. As the relaxation of hot electrons accelerates with smaller electronic specific heat, and because $c_v^e$ is proportional to the density of states at the Fermi level [3], for metals relaxation should *accelerate* linearly with the inverse of the bandwidth $\tau \propto 1/t$.

For solids with strongly correlated electrons, the dependence of non-equilibrium quasiparticle recombination rates on the microscopic parameters has not been systematically investigated and it is not well understood. In this letter, we study the recombination of impulsively excited quasi-particles in a one dimensional Mott insulator, in which we tune electronic bandwidth and intersite correlation energy by applying external pressure. We find that the recombination of quasi-particles accelerates for increasing bandwidth, as expected for a metal, but with a dependence on microscopic parameters that is unique to the physics of electronic insulators in one dimension and that descends from a competition between local decay and coherent delocalization of photo-excited holon-doublon pairs [4].

We study bis-(ethylendithyo)-tetrathiafulvalene-difluorotetracyano-quinodimethane (ET-F$_2$TCNQ), a half filled organic salt with quasi-one dimensional electronic structure, negligible electron-phonon interaction [5] and with electronic properties that are well captured by a 1D extended Hubbard model [4,6]

$$H = -t\sum_{j\sigma}\left(\hat{c}_{j\sigma}^{+}\hat{c}_{j+1\sigma} + h.c.\right) + V\sum_j \hat{n}_j\hat{n}_{j+1} + U\sum_j \hat{n}_{j\uparrow}\hat{n}_{j\downarrow}. \tag{1}$$

Our experiments were performed under external pressure, which reduces the lattice spacing between the molecular sites and tunes the hopping amplitude ($t$) and repulsion (attraction) between electron (electron-hole) pairs on neighboring sites ($V$). In the simplified case of hydrogenic wavefunctions in one dimension [7,8], $t$ depends exponentially on the lattice spacing $R$ as $t(R) \propto (1+\alpha R)e^{-\alpha R}$. The decrease in the lattice spacing $R$ results then in an increase of the intersite Coulomb repulsion $V$ as $V(R) \propto e^2/R$ [9]. Within these simplifying assumptions, the onsite Coulomb interaction $U$, which is determined by the local electronic properties of the molecular sites, is considered independent on the lattice spacing [10].

For calibration, we first measured and fitted the static optical properties as a function of pressure (see supplementary material). In Fig. 1, we report the reflectivity measured at 300 K for electric field polarized along the 1D charge transfer direction (*a*-axis). The corresponding optical conductivity $\sigma_1(\omega)$ was extracted by a Kramers-Kronig consistent fit. At ambient conditions (see Fig. 1(a) and (b)), $\sigma_1(\omega)$ exhibits a prominent peak near 700 meV, indicative of electron-electron correlations and of a Mott gap [5]. As pressure was applied, the 700 meV Charge Transfer (CT) resonance was observed to shift to the red at a rate of 70 meV/GPa, broadening toward high frequencies (see Fig. 1(c) and (d)). This pressure dependence is consistent with the behavior measured previously in other quasi-1D compounds [11,12]. For all measured photon energies (>75 meV), no Drude response was observed at any pressure (0 - 2 GPa), indicating that the material remains insulating and one-dimensional [12,13]. The vibrational peaks at frequencies below 400 meV exhibited no pressure dependence, excluding significant intramolecular structural rearrangement.

The data were analyzed using a model of the optical conductivity based on the extended Hubbard Model of Eq. (1). The onsite repulsion $U$ was assumed to be a constant 845 meV at all pressures. The large Mott gap of ET-F$_2$TCNQ enables a $1/U$ strong coupling expansion of the extended Hubbard model giving a reduced optical conductivity $\omega\sigma_1(\omega)$ in the form

$$\omega\sigma_1(\omega) = g_0 t^2 e^2 \left\{ \Theta(V-2t)\pi\left[1-\left(\frac{2t}{V}\right)^2\right]\delta(\omega-\omega_{CT}) + \Theta(4t-|\omega-U|)\frac{2t\sqrt{1-\left[\frac{\omega-U}{4t}\right]^2}}{V(\omega-\omega_{CT})} \right\} \qquad (2)$$

where $e$ is the electronic charge, $\Theta$ is the Heaviside function and $g_0 = 2.65$ is the zero-momentum form factor accounting for the spin degrees of freedom [14,15] (see supplemental material).

This analytical result highlights the two dominant contributions to $\omega\sigma_1(\omega)$ resulting from the relevant quasi-particles of this system. The first is a delta peak located at $\omega_{CT} = U - V - 4\,t^2/V$, and corresponds to a Mott-Hubbard exciton composed of a bound holon-doublon (HD) pair. A second broad peak is centered around $U$, with a bandwidth of $8t$, and corresponds to the continuum of states associated with unbound particle-hole (PH) excitations [14,15]. These two types of excitations are visualized for the 1D lattice model in Fig. 2(d). As pressure is applied, the exciton peak shifts to the red, whereas the continuum remains centered at $U$ and broadens with the increase in bandwidth.

A comparison of the measured and fitted $\omega\sigma_1(\omega)$, normalized to the peak value of the spectrum measured at ambient pressure, is reported for three selected pressures in Fig. 2(a) - (c) as black (measured data) and grey (theory fits) curves. The fit has been performed here after subtraction of the low-frequency tail of the intramolecular absorption bands near 3.5 eV [4,5]. The fitted values for $t$ and $V$ are summarized in Fig. 2(e). The nearest neighbor interaction $V$ and the hopping amplitude $t$ are both observed to increase with pressure, from 120 meV to 203 meV and from 40 to 85 meV, respectively.

Optical pump-probe experiments were then used to measure the ultrafast recombination after prompt photo-excitation of holon-doublon pairs. An amplified Ti:Sa laser, operating at 1 KHz repetition rate and generating 800-nm-wavelength pulses, was used to drive a near infrared optical parametric amplifier, which was used for degenerate pump-probe experiments ($\lambda_{pump} = \lambda_{probe}$) in a diamond anvil cell.

In a first series of experiments, the 100 fs pulses were tuned between 1.6 μm and 2.2 μm wavelengths, to track the peak of the conductivity at each pressure value (see Fig. 2) and thus excite bound holon-doublon pairs. Above 1.3 GPa the tuning was no longer possible, as for higher pressures the laser spectrum overlapped with the multiphonon absorption in the diamond anvil.

A second set of experiments was performed with pump and probe wavelengths fixed on the center of the pressure-independent particle-hole continuum band (740 meV, 1.7 μm wavelength). Both experiments were performed with a pump fluence of 300 μJ/cm², in a regime in which the excitation of holon-doublon pairs is sparse and the signal scales linearly with the number of absorbed photons.

In Fig. 3 the normalized time-resolved reflectivity changes ΔR/R are reported on a semi-logarithmic scale (see panels (a) and (c)). In both cases, photo-excitation promptly reduced the holon-doublon band and the particle hole continuum, with a drop in reflectivity ΔR/R ~ -1%, followed by a rapid recovery of the signal with a double exponential law [4,6]. Only the fast time constant showed pressure dependence, with the recombination rate decreasing at a rate of 116 fs/GPa (490 fs to 350 fs between 0 and 1.3 GPa) for bound holon-doublon pairs and 85 fs/GPa (470 fs to 300 fs between 0 and 2 GPa) for the particle hole continuum. Both holon-doublon and particle-hole excitations exhibited similar recombination times and similar dependence on pressure. Indeed, from the static fitting discussed above, the ratio $V/t$ is reduced from $V/t \sim 3.0$ to $V/t \sim 2.4$, close to the critical ratio $V/t \sim 2$ for which the exciton peak cannot be isolated from the continuum [14,15]. The slow time constant of 2.5 ps, interpreted here as resulting from the thermalization of high-energy molecular modes heated by the recombination of quasi-particles, was independent on pressure.

To analyze this pressure dependence quantitatively, we first note that recombination of bound holon-doublon pairs (or disassociated pairs throughout the particle hole continuum) must involve dissipation of an excess energy of order $U - V$ by coupling to a bath. In absence of such bath the only possible decay paths involve kinetic energy transfer (creation of particle-hole pairs) or spin excitations within the Hubbard model itself, where the relevant energy scales are the bandwidth $t$ and the exchange coupling $t^2/(U - V)$, respectively [16]. For $U \gg V, t$ a large number of scattering events is required for recombination, leading to a decay rate that is exponentially small with $U/t$ [16,17,18,19]. Owing to the weak electron-phonon coupling in ET-F$_2$TCNQ [5], and lack of efficient radiative emission, it is reasonable to assume that high-frequency molecular modes are the primary scattering partner for the rapid decay of hot quasi-particles observed in our experiment.

In the following discussion, we assume that recombination occurs locally and the density of holon-doublon excitations is sufficiently low that only a single pair need be considered. This is substantiated by the measured fluence dependence. In the range explored in our experiments the signal scaled linearly with the laser fluence, while recombination rates were found to be independent of it. This behavior suggests that excitations are sparsely distributed and recombination is localized.

The simplest possible model involves a dimer of ET molecules coupled to a continuum of bosonic modes, as is commonly used to model dissipative electron transfer [20,21]. The holon-doublon and singly-occupied configurations of the dimer, with an energy gap $U - V$ and tunneling $t$ between them, comprise a two-level system of the spin-boson model [22]. We take a super-ohmic form for spectral function $\mathcal{J}(\omega)$ appropriate for electron-phonon interactions [23]. Moreover, given $U - V \gg t$, the system is in the large-bias regime where the far-off-resonant tunneling mediates the decay of the holon-doublon pair by dissipating the large energy gap into the bosonic reservoir. The rate of decay $\Gamma$ in this limit is proportional to $\left(\frac{t}{U-V}\right)^2 \mathcal{J}(U-V)$ [22]. As the dominant contributions to the spectral function at the energy gap $\mathcal{J}(U-V)$ arise from local vibrational modes it is assumed to be independent of pressure. This then gives a scaling of $\Gamma$ with the pressure dependent microscopic parameters as $\left(\frac{t}{U-V}\right)^2$, which sensibly predicts an increase with $t$ and a decrease with $U - V$.

In Fig. 4 the experimentally measured holon-doublon and particle-hole continuum decay rates, normalized to the ambient pressure decay rate $1/\tau_0$, are plotted as a function of $(U-V)/t$. The dashed line shows the predicted change in the recombination rate (see supplemental material for details), consisting of a more than five-fold increase at the largest pressure. Thus, the two-site model on its own does not reproduce the experimentally measured trend.

We note that previous time resolved spectroscopy experiments showed that at least three sites are necessary to account for the photo-response [4], owing to the possibility for bound holon-doublon pairs next to a singly occupied site $|\cdots \_ 0 \updownarrow \_ \_ \cdots \rangle$ to tunnel into a configuration of the type $|\cdots \_ 0 \_ \updownarrow \_ \cdots \rangle$, where $|\cdots 0 \cdots \rangle$ represents a holon, $|\cdots \updownarrow \cdots \rangle$ a doublon and $|\cdots \_ \cdots \rangle$ represents a singly occupied site of either spin. This separation occurs at a rate determined by the hopping amplitude $t$ and is limited by a barrier $V$, where both of these increase with pressure. Importantly, the $|\cdots \_ 0 \_ \updownarrow \_ \cdots \rangle$ configuration has a lower probability to recombine than the configuration $|\cdots \_ 0 \updownarrow \_ \cdots \rangle$, as two hopping events are necessary.

To include this effect, we employed an effective model based on the sketch of Fig. 5, derivable from the strong-coupling limit (see supplemental material), where in addition to the singly-occupied configuration $|g\rangle$ and adjacent holon-doublon pair $|0\rangle$, other states $|l\rangle$, representing holon and doublons separated by $l$ sites were considered, up to maximum distance $L$. The potential energy of the model then mimicked the interactions of the equivalent many-body configurations in the full, extended Hubbard model. Importantly, the optical conductivity of this effective model, when $L \to \infty$, exactly reproduces that of Eq. (2) for the extended Hubbard model (see supplemental material).

Local recombination was added to this effective model via a Markovian quantum dissipation process, appropriate for the super-ohmic large-bias limit [24] (see supplemental material), that incoherently drives only the transition $|0\rangle \to |g\rangle$ at a bare rate $\Gamma$. This effective model reveals that even the addition of one ionized holon-doublon state $|1\rangle$ which is unaffected by the decay, i.e. taking $L = 1$, causes the suppression of the actual decay rate $\Gamma_{eff}$ to $|g\rangle$ from the bare rate $\Gamma$ as

$$\Gamma_{eff} = \frac{\Gamma}{2}\left(1 + \frac{V}{\sqrt{V^2 + 16t^2}}\right). \tag{3}$$

We applied this effective model by first fitting the bare rate $\Gamma$ to reproduce the observed zero pressure decay rate. For all other pressures $\Gamma$ was increased according to the spin-boson scaling by using the values for $t, V$ reported in Fig. 2(e). The effective decay rate $\Gamma_{eff}$ was then calculated numerically from this model for different sizes $L$. As $L$ increases the suppression of $\Gamma_{eff}$ becomes more pronounced since disassociated holon-doublon pair can ballistically separate to larger distances remaining immune to the local decay. Thus, we find that coherent dynamics on more than two sites, accounting for the break-up of the holon-doublon pair, competes and slows down the recombination caused by the local decay process.

In conclusion, we have experimentally investigated the pressure dependence of hot quasi-particle recombination in a one dimensional Mott insulator. By fitting the steady state infrared properties with a model based on the extended Hubbard Hamiltonian, we extracted the pressure dependence of the Hubbard parameters $t$ and $V$ up to 2.0 GPa, and correlated them to the recombination rates. A key inference made by comparing the experimentally determined dependence to theory is that the decay of quasi-particles is likely connected to the coherent evolution of holon-doublon pairs immediately after excitation. Based on this idea, it may be possible in the future to accelerate or decelerate the photo-induced dynamics of correlated electron systems by pulse-shaping and by coherent optical control techniques. Such ability would have interesting ramifications in both fundamental science and applications.

Acknowledgments

We thank P. Di Pietro and A. Dengl for technical support in the equilibrium optical measurements and M. Eckstein for fruitful discussions. This research has been funded by the European Research Council under the European Union's Seventh Framework Programme (FP7/2007-2013) / ERC Grant Agreement n° 319286. D.J. and S.R.C. thank the National Research Foundation and the Ministry of Education of Singapore for support.

**FIGURES**

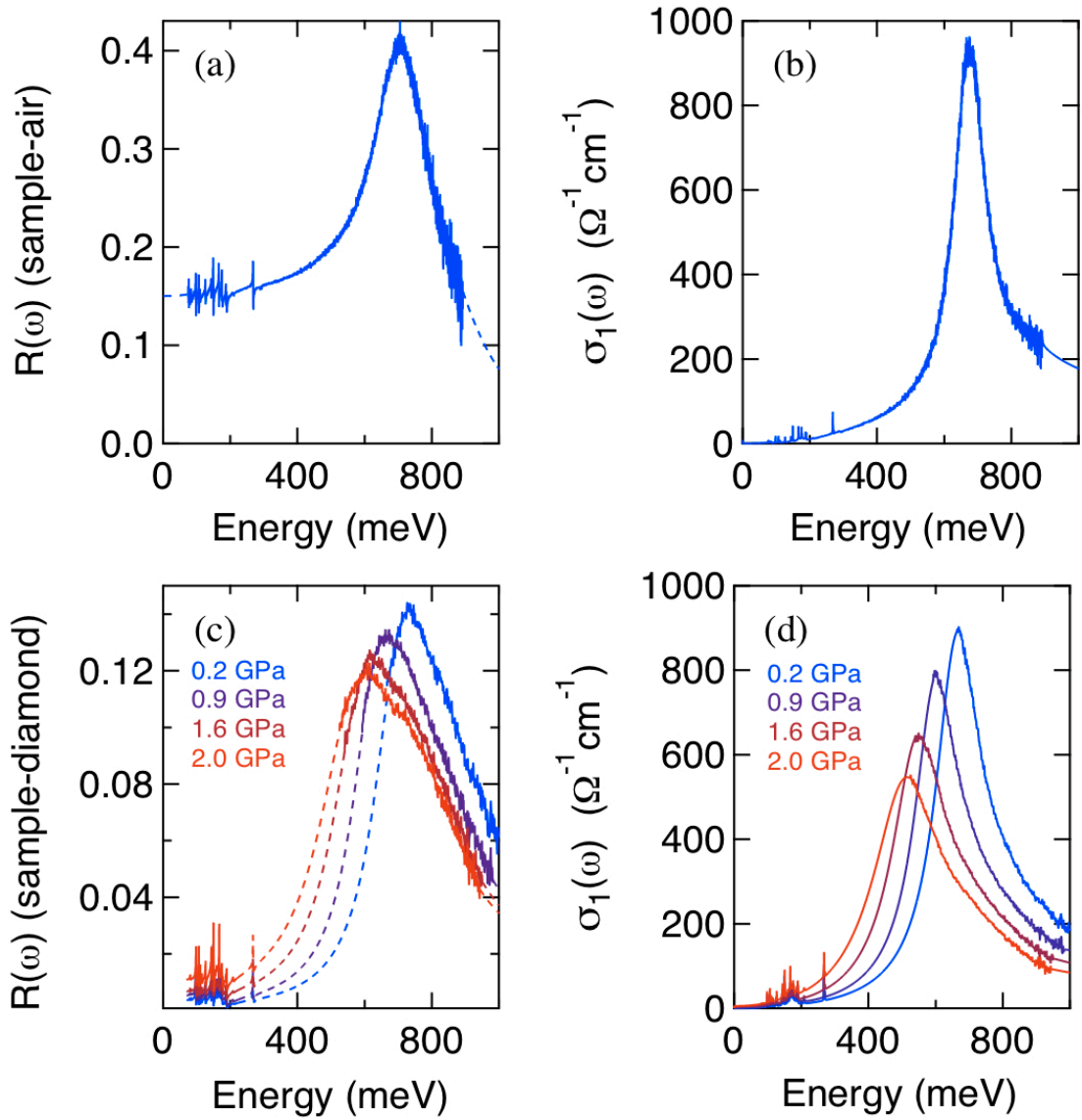

Fig. 1 (Color online) (a) Static reflectivity at ambient pressure of the ET-F$_2$TCNQ, measured with the electric field parallel to the *a* axis. (b) Real part of the optical conductivity $\sigma_1(\omega)$ at ambient pressure. (c) Steady state reflectivity of the ET-F$_2$TCNQ along the *a* axis for selected pressures. (d) Pressure dependence of the optical conductivity $\sigma_1(\omega)$.

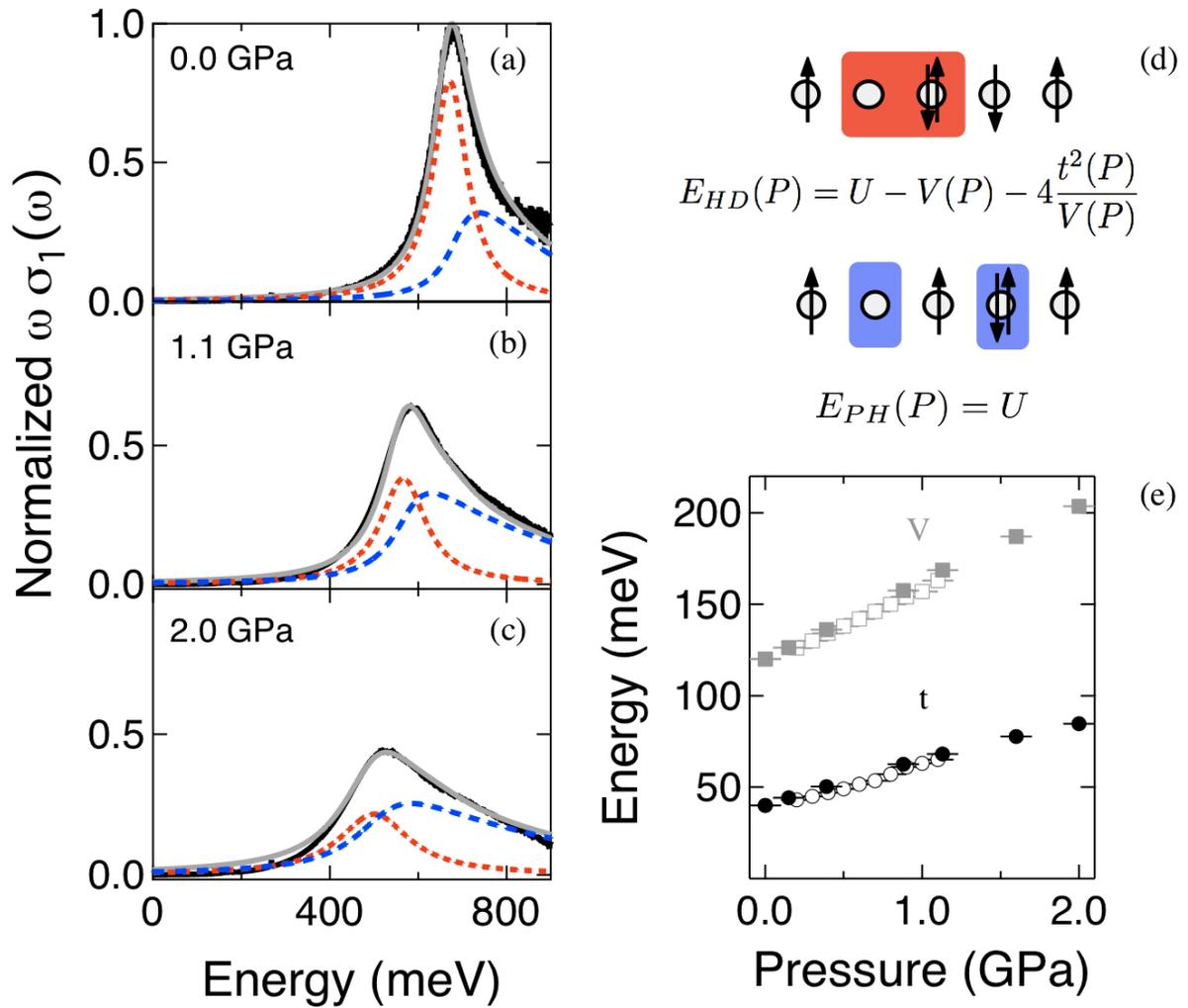

Fig. 2 (Color online) (a)-(c) Normalized reduced optical conductivity for selected pressures (black solid line), with corresponding fit (grey solid line). The contributions of the holon-doublon pair (red) and of the particle-hole continuum (blue) are shown as dashed lines. (d) Sketch of a holon-doublon pair (top) and of a typical particle-hole continuum excitation (bottom). (e) Pressure dependence of the extended Hubbard model parameters V (squares) and t (circles) extracted from the fit of the steady state $\omega \sigma_1(\omega)$. U is kept fixed to 845 meV. Filled and empty symbols identify distinct experimental runs.

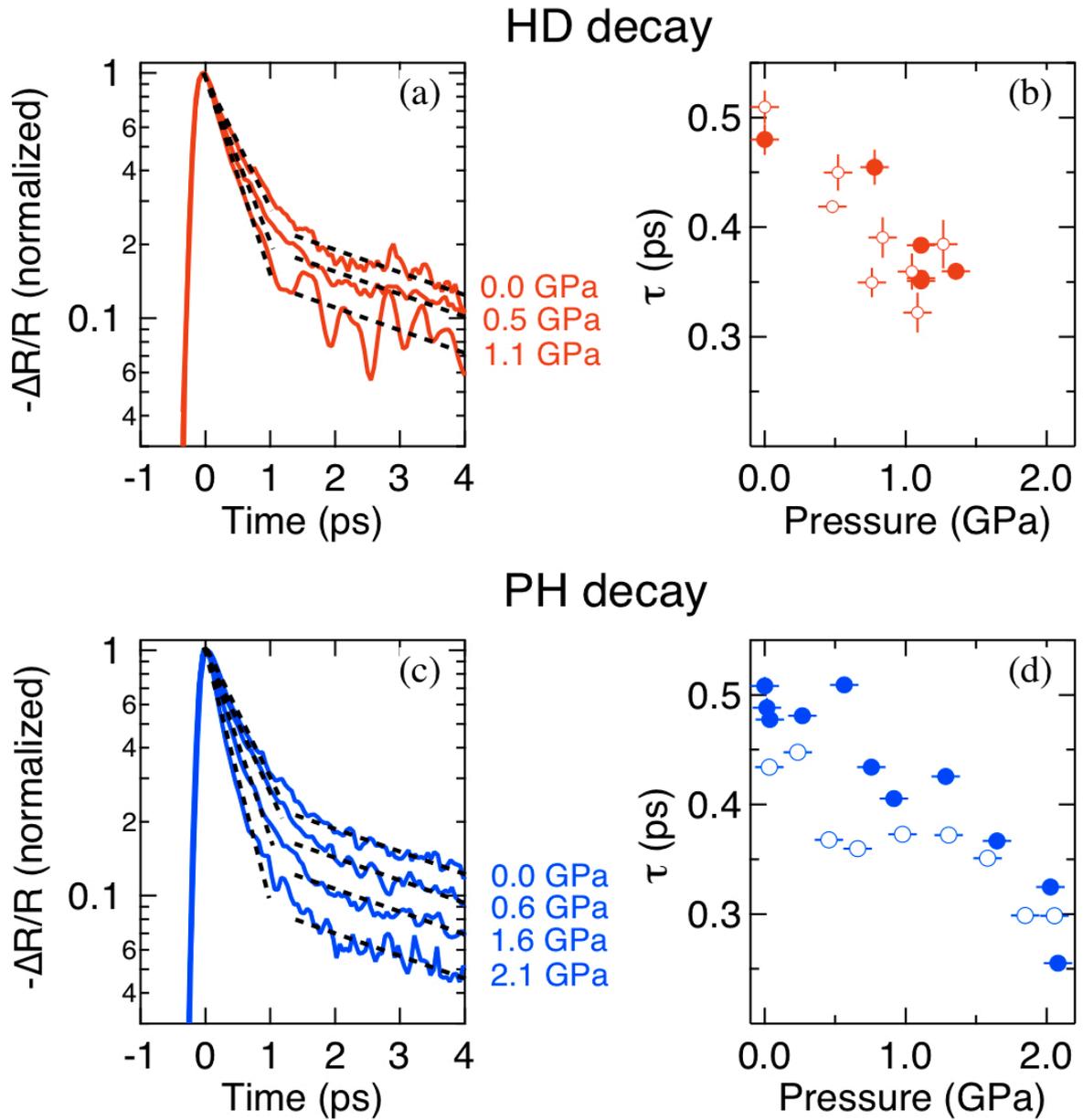

Fig. 3 (Color online) (a) Normalized $\Delta R/R$ time domain curves on the holon-doublon pair peak (HD) for selected pressures (solid lines). Black dashed lines are guides to the eye showing the bi-exponential decay, linear on a log scale. (b) Holon-doublon recombination lifetimes extracted from a fit to the data. Filled and empty symbols identify distinct experimental runs. (c) Normalized $\Delta R/R$ time domain curves on the particle-hole continuum peak (PH) for selected pressures (solid lines). Black dashed lines are guides to the eyes showing the bi-exponential decay. (d) Particle-hole continuum recombination lifetimes extracted from a fit to the data. Filled and empty symbols identify distinct experimental runs.

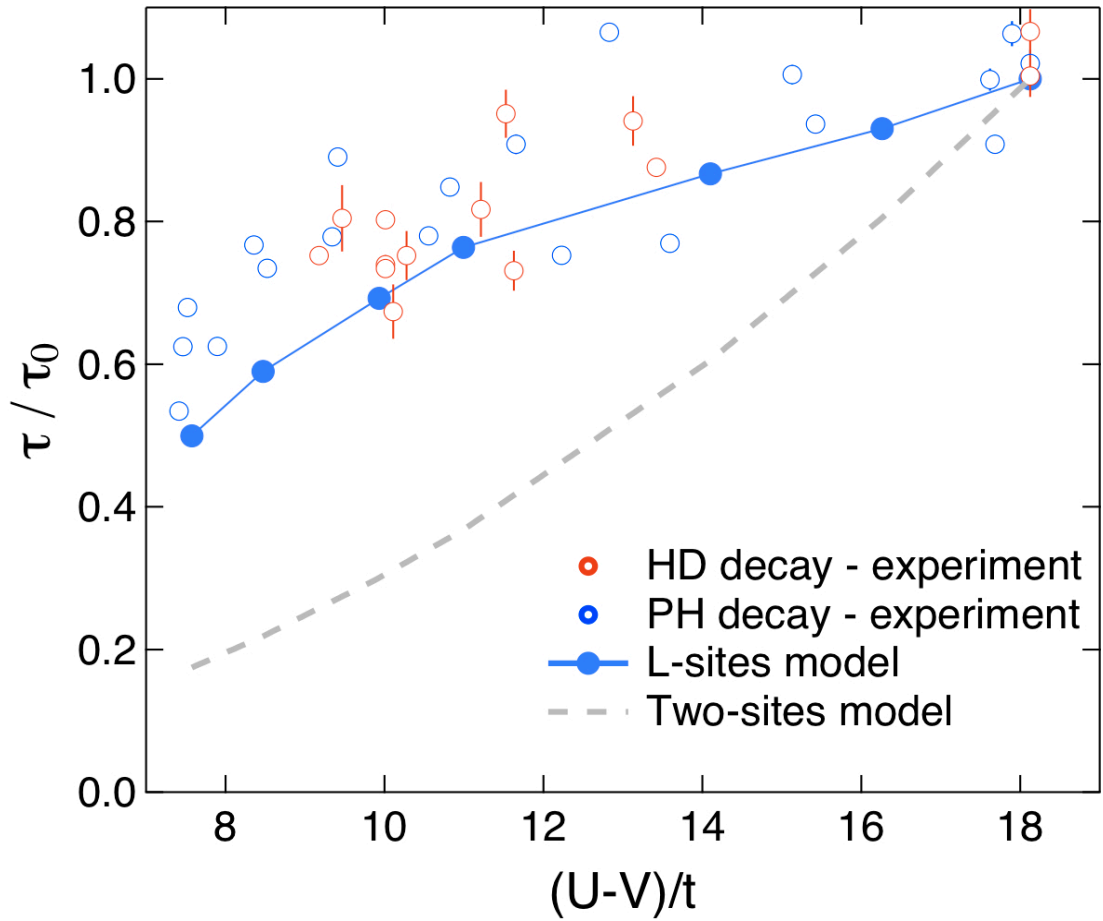

Fig. 4 (Color online) Experimental relaxation times (empty circles) for both HD and PH resonant excitation normalized to their ambient pressure values $\tau_0$ and shown as a function of $(U-V)/t$. The calculated holon-doublon lifetimes (filled circles) found from solving an $L$-site effective model (see Fig. 5) with spin-boson scaling for the bare decay rate. For comparison the $L = 1$ dimer result is shown (dashed line).

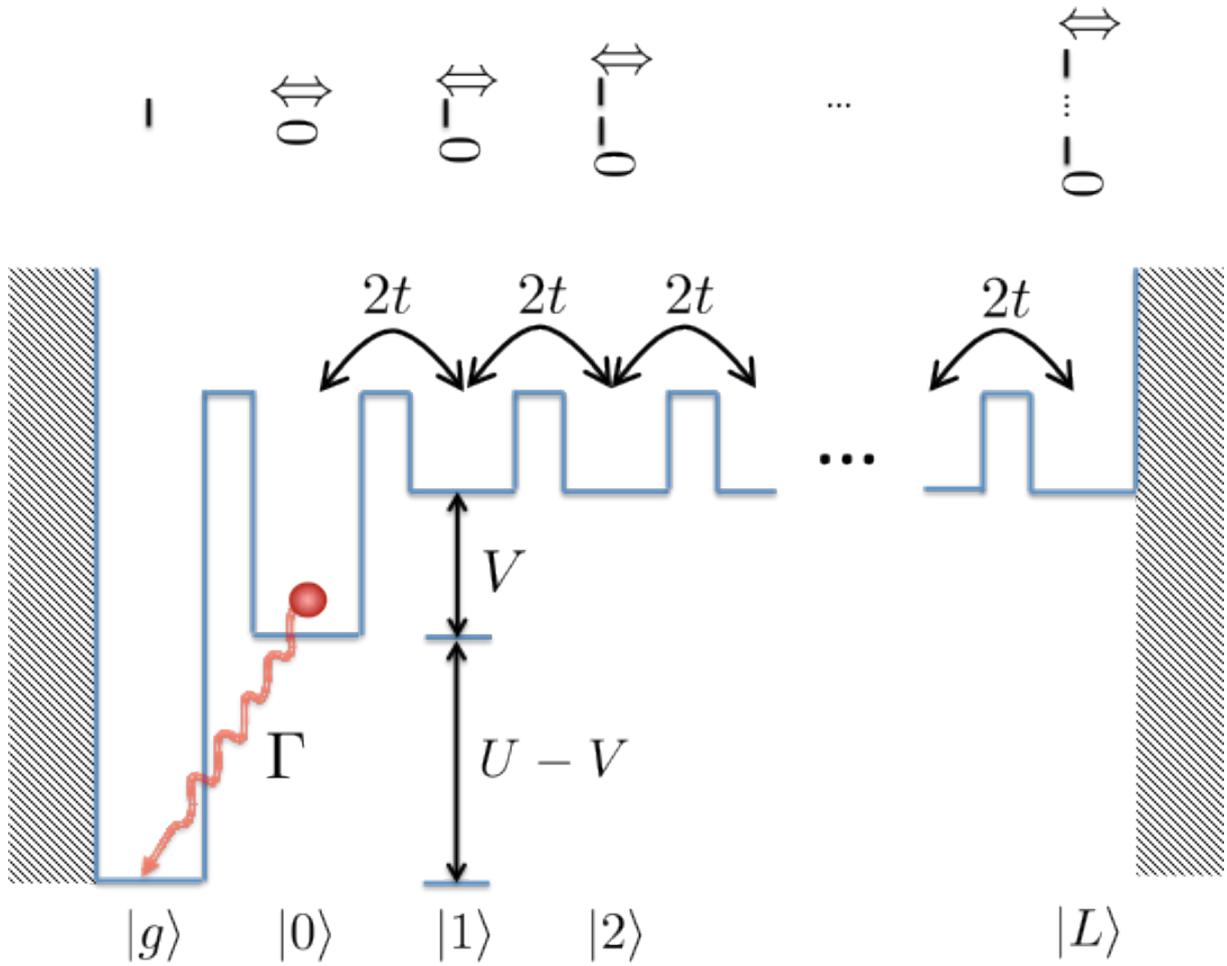

Fig. 5 (Color online) Effective model describing holon-doublon dynamics in the strong-coupling limit. The state $|g\rangle$ is the ground state containing no holons or doublons, whereas the state $|0\rangle$ represents an adjacent (zero separation) holon-doublon pair. The remaining states $|l\rangle$ represent the holon and doublon being separated by $l$ sites. In the limit $L \to \infty$ these unbound states form the particle-hole continuum. The relaxation to $|g\rangle$ at a bare spin-boson rate $\Gamma$ only occurs locally from state $|0\rangle$.

# Supplemental Material:
# Pressure dependent relaxation in the photo-excited Mott insulator ET-F$_2$TCNQ: influence of hopping and correlations on quasi-particle recombination rates

M. Mitrano[(1)], G. Cotugno[(1,2)], S.R. Clark[(3,2)], R. Singla[(1)], S. Kaiser[(1)], J. Staehler[(4)], R. Beyer[(5)], M. Dressel[(5)], L. Baldassarre[(6)], D. Nicoletti[(1)], A. Perucchi[(7)], T. Hasegawa[(8)], H. Okamoto[(9)], D. Jaksch[(2,3)], A. Cavalleri[(1,2)]

[(1)]*Max Planck Institute for the Structure and Dynamics of Matter, Luruper Chaussee 149, Hamburg, Germany*

[(2)] *Department of Physics, Oxford University, Clarendon Laboratory, Parks Road, Oxford, United Kingdom*

[(3)]*Centre for Quantum Technologies, National University of Singapore, Singapore*

[(4)]*Fritz Haber Institute of the Max Planck Society,* Faradayweg 4-6, 14195 Berlin, Germany

[(5)]*1. Physikalisches Insitut, Universität Stuttgart, Pfaffenwaldring 57, 70550 Stuttgart, Germany*

[(6)] *Center for Life NanoScience@Sapienza, Istituto Italiano di Tecnologia, V.le Regina Elena 291, Rome, Italy*

[(7)] *INSTM UdR Trieste-ST and Elettra - Sincrotrone Trieste S.C.p.A., S.S. 14 km 163.5 in Area Science Park, 34012 Basovizza, Trieste Italy*

[(8)]*National Institute of advanced Industrial Science and Technology, Tsukuba, Japan*

[(9)]*Department of Advanced material Science, University of Tokyo, Chiba 277-8561, Japan*

## A. Determination of the steady state optical constants under pressure

1) Experimental details

The equilibrium infrared measurements were performed with two different setups, one located in the University of Stuttgart (Germany), the other in the beamline SISSI at the Elettra Synchrotron Radiation Facility of Trieste (Italy). All the measurements were performed at room temperature and in quasi-normal incidence conditions with linearly polarized light along the crystalline $a$ axis.

In the first setup, infrared reflectivity spectra up to 1.1 GPa were collected with a piston-cylinder cell coupled to a Bruker IFS66v interferometer (diamond window diameter 3 mm). The hydrostatic medium chosen in that case was Daphne 7373 oil, and internal pressure has been monitored through the external applied pressure. In order to have a reliable indication of the internal pressure a preliminary calibration with the ruby fluorescence method [1] has been performed.

The second setup was based on a screw-diamond anvil cell (600 μm culet size) coupled to a Bruker Vertex70 interferometer through a Hyperion microscope equipped with Schwarzchild objectives [2]. Here both reflectivity and transmittance spectra were collected for each pressure point up to 2.0 GPa. The hydrostatic medium chosen for these measurements was CsI powder. Pressure was measured in-situ with the standard ruby fluorescence method [1]. Details about the referencing of the spectra are given in Ref. [2].

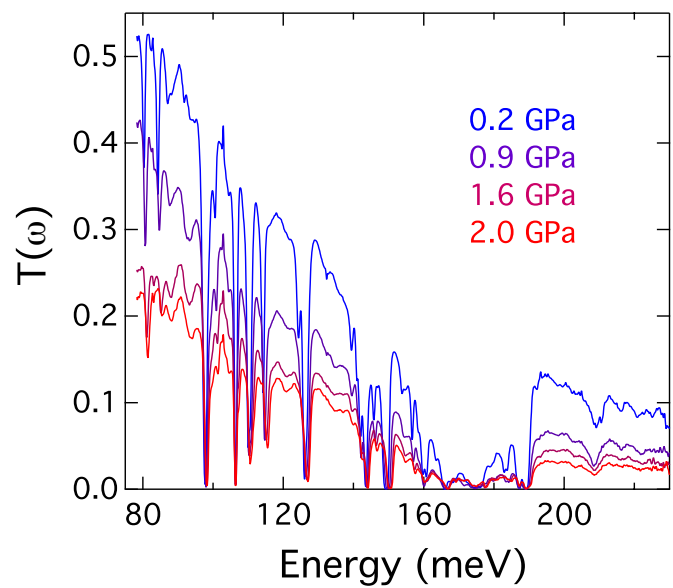

Fig. S1 – Steady state mid-infrared transmittance of ET-F$_2$TCNQ along the a axis for selected pressures.

2) Mid-infrared Transmittance

In order to reliably extract the optical conductivity of the ET-$F_2$TCNQ samples, we complement our reflectivity measurements by following the pressure dependence of the transmittance of a thin (10 μm) slab of sample in the mid-infrared region. Figure S1 shows the transmittance data for selected pressures. For energies above 240 meV the presence of the diamond anvil absorption prevents a reliable measurement of the transmittance.

3) Determination of the optical conductivity

In a diamond anvil cell the measurement of the reflectivity in the mid-infrared region is challenging because of the similarity of the refractive indices of the sample (n=2.60) and of the diamond anvil (n=2.37). Therefore the determination of the pressure dependent optical conductivities for the ET-$F_2$TCNQ is performed through a simultaneous Drude-Lorentz (DL) fit [2] of both the reflectivity and transmittance data for each pressure point.

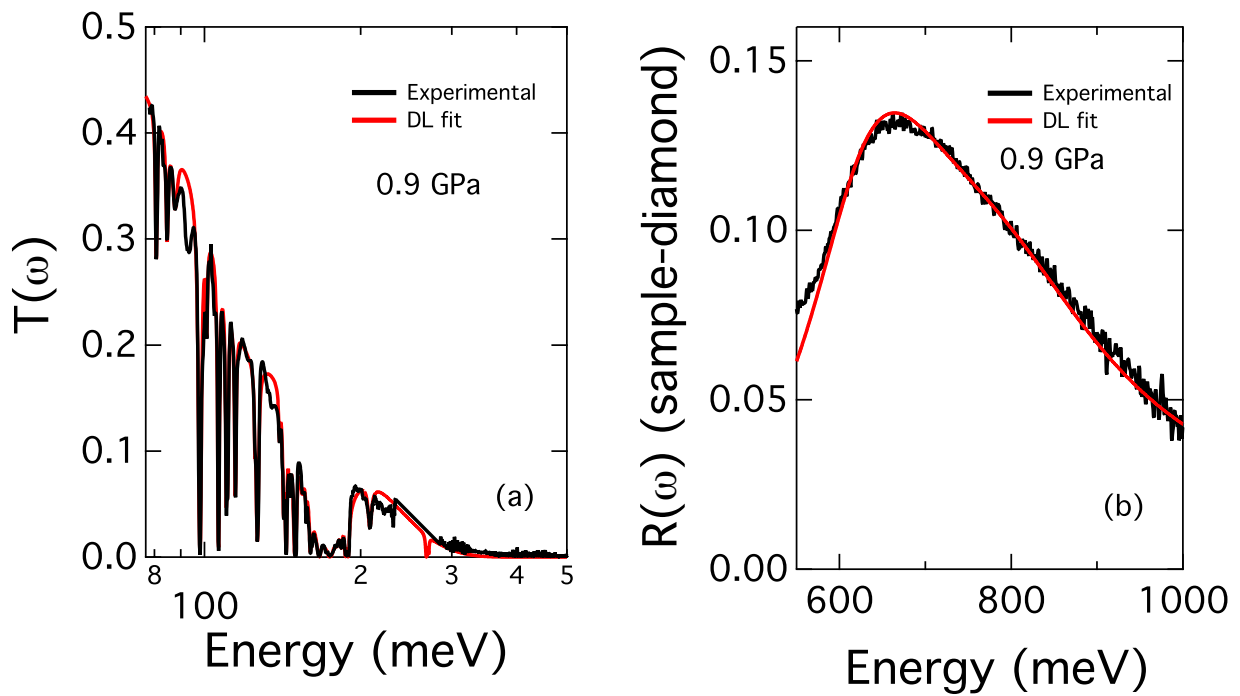

Fig. S2 – Sample Drude-Lorentz (DL) fit for the steady state mid-infrared transmittance and near-infrared reflectivity of the ET-$F_2$TCNQ along the a axis (at 0.9 GPa).

In detail, a Kramers-Kronig consistent DL fit is made on the transmittance for energies up to 300 meV, while for higher energies (up to 1000 meV) the reflectivity is fitted (see Fig. S2).

From the DL fit of the transmittance data, the low frequency part of the bulk reflectivity is then recalculated at the sample-diamond interface and merged with the experimental data above the multiphonon diamond absorption (namely above 550 meV).

The obtained full-range reflectivity at the sample-diamond interface is then used to calculate the optical conductivity $\sigma_1(\omega)$ through a standard Kramers-Kronig algorithm for samples in contact with a transmittive window [3]. The $\beta$ parameter for the Kramers-Kronig transformation is kept fixed at 471 meV by checking the consistency of the output $\sigma_1(\omega)$ with the ambient pressure data where the Kramers-Kronig procedure is independent from this parameter. The merged reflectivities and the corresponding optical conductivities are reported in Fig. 1(c) – (d) of the main text.

**B. Steady state theory fit of the optical conductivity**

In the limit of large $U \gg t, V$ it is possible to analyze the extended Hubbard model by means of a $1/U$ expansion [4]. If we ignore corrections of order $t/U$ then electron transfers are limited to those which conserve the number of doublons and in the ground state all sites are singly occupied. In this limit spin and charge dynamics decouple, leading to spin-charge separation, and a simple upper and lower Hubbard band picture emerges for the charges. The spin-dependence of the optical conductivity then enters only via a momentum dependent ground-state spin correlation $g_q$. This is further simplified by using the so-called "no-recoil approximation" where the dominant contributions to the conductivity are argued to arise from $q = 0$ and $q = \pi$ transitions [4], where $g_0 = 2.65$ and $g_\pi = 0.05$. The reduced optical conductivity is then given by

$$\omega\sigma_1(\omega) = \pi g_\pi t^2 e^2 \delta(\omega - \omega_2)$$

$$+ g_0 t^2 e^2 \left\{ \Theta(V-2t)\pi\left(1 - \frac{4t^2}{V^2}\right)\delta(\omega - \omega_1) \right.$$

$$\left. + \Theta(4t - |\omega - U|) \frac{2t\sqrt{1 - (\omega - U/4t)^2}}{V(\omega - \omega_1)} \right\},$$

where $\Theta(x)$ is the Heaviside step function, $\omega_1 = U - V - 4t^2/V$ is the exciton energy and $\omega_2 = U - V$. The two $\delta$ peaks correspond to two different Mott-Hubbard excitons, with the $\Theta(V - 2t)$ factor expressing that the $\omega_1$ exciton exists only when $V > 2t$. Owing to the small value of $g_\pi$ the $\omega_2$ exciton was ignored in our analysis due to its negligible weight. The final term is a semi-elliptic contribution from the particle-hole continuum which exists only for $|\omega - U| \leq 4t$. To allow comparison to the experiment the sharp features arising from the $\delta$ and $\Theta$ functions where convolved with a Lorentzian inducing a broadening $\eta \sim 2t$. The subsequent best-fit focused on the features near the CT resonance and provided the changes in the microscopic parameters displayed in Fig. 2(e) of the main text.

### C. High-energy spinless fermion model

In the strong coupling limit it's possible to obtain an a simple effective Hamiltonian that describes the high-energy physics associated with particle-hole excitations across the Mott gap. This is achieved by first splitting the hopping terms into parts, one which conserves the number of double occupancies and one which does not. Following a well-known procedure [5], one seeks a canonical transformation $S$, which removes the order $t$ non-conserving contributions. Retaining only the lowest order terms in $t$ after this transformation then yields an effective Hamiltonian where the number of double occupations is a good quantum number. The non-conserving terms are now of order $\sim t^2/U$ and can be neglected in the strong-coupling limit $U \gg t$.

Exact eigenstates of this effective model can be found [6]. This is accomplished by defining spinless fermionic operators $\hat{a}_j$ for each site $j$ which act only on the occupation indices $\vec{n} \equiv (n_1, \ldots, n_N)$, where

$n_j = 0$ or 1 counts the number of particles at site $j$. Charge excitations of the system can then be shown to be described by a Hamiltonian

$$H_{exc} = (U - V) \sum_j \hat{n}_j \hat{n}_{j+1} - t \sum_j (\hat{a}_j^+ \hat{a}_{j+1} + \hat{a}_{j+1}^+ \hat{a}_j),$$

where $\hat{n}_j = \hat{a}_j^+ \hat{a}_j$ is the number operator. In the subspace of a single holon-doublon pair a convenient basis for the description of these charge excitations (or excitons) is in terms of a centre-of-mass coordinate $R$, and a relative coordinate $r$. This has the form

$$|\Psi_{HD}\rangle = \sum_{rR} \Phi(r, R) |R + r/2, R - r/2\rangle,$$

where $\left|R + \frac{r}{2}, R - \frac{r}{2}\right\rangle = \hat{a}_{R+r/2}^+ \hat{a}_{R-r/2}^+ |0\rangle$ with $|0\rangle$ being the vacuum of the two-body problem. The exciton is a two-particle bound state, so one looks for a solution in analogy to the hydrogen atom. The scalar product $\langle R + \frac{r}{2}, R - \frac{r}{2}|H_{exc}|\Psi_{HD}\rangle$ gives a difference equation for the exciton wavefunction [7]. Without loss of generality we can take the centre-of-mass momentum $K$ to be $K = 0$ and the Schrödinger equation becomes

$$2t[2\psi_n(r) - \psi_n(r-a) - \psi_n(r+a)] - 4t\psi_n(r) - V\delta_{ra} + U\psi_n(r) = E\psi_n(r).$$

This is equivalent to single particle effective Hamiltonian defined on a semi-infinite tight-binding chain

$$H_{hd} = (U - V)|0\rangle\langle 0| + U \sum_{l=1}^{L} |l\rangle\langle l| - 2t \sum_{l=0}^{L-1} (|l\rangle\langle l+1| + |l+1\rangle\langle l|).$$

In Fig. 5 of the main text this effective model is depicted for $L$ sites with open boundary conditions. The ground state configuration $|g\rangle$ signifying no holon-doublon pair has zero potential energy, while the adjacent holon-doublon configuration $|0\rangle$ has a potential energy $U - V$, and all the more distant holon-doublon configurations $|l\rangle$, where $l = 1, 2, 3, \cdots L$ have a potential energy $U$. The hopping between neighbouring holon-doublon configurations, e.g. $|l - 1\rangle$ and $|l\rangle$, is included, but no hopping occurs between $|g\rangle$ and $|0\rangle$ because the coherent evolution in this strong-coupling limit conserves the holon and doublon. Note that the hopping amplitude is $2t$ because the model describes the relative motion and the reduced effective mass of holon-doublon pair is half that of either excitation individually.

The current operator in this picture reduces to $J \propto |g\rangle\langle 0| + |0\rangle\langle g|$ which mimics the optical excitation of the Mott insulating ground state $|g\rangle$ through the creation of an adjacent (bound) holon-doublon configuration $|0\rangle$. The optical conductivity of the model $\sigma_1(z) = \langle g|J(z-H)^{-1}J|g\rangle$ is then equivalent to the single-particle Green function $G(z) = \langle 0|(z-H)^{-1}|0\rangle$, with complex frequency $z$. In the limit $L \to \infty$ this is found to be

$$G(z) = \frac{2}{z - U + 2V \pm \sqrt{(z-U)^2 - (4t)^2}}.$$

Expanding the imaginary part of $\omega\, G(\omega)$ and extracting the residue of the poles reproduces the dominant $\omega_1$ Mott-Hubbard exciton $\delta$ peak and the particle-hole continuum found from the strong-coupling result in Eq. (2) in the main text. In other words, apart from the spin-averaging numerical factor $g_0$, the same spectral features of the extended Hubbard model in the limit $U \gg V, t$ are in fact captured by this simple effective model. Crucially the $2t$ hopping amplitude for the relative motion is responsible for the $8t$ bandwidth of the particle-hole continuum.

### D. Spin-boson scaling of holon-doublon decay

To model holon-doublon recombination we apply a generic, but well-established description based on the celebrated spin-boson model [8]. Specifically, we consider focus on the ground state $|g\rangle$ and adjacent holon-doublon configuration $|0\rangle$ as a two-level system separated in energy by $U - V$. Then by coupling the electron density linearly to a continuum of bosonic modes the following spin-boson Hamiltonian is obtained

$$H_F = -t(|0\rangle\langle g| + |g\rangle\langle 0|) + (U-V)|0\rangle\langle 0| + \sum_n \lambda_n(\hat{b}_n^+ + \hat{b}_n)|0\rangle\langle 0| + \sum_n \omega_n \hat{b}_n^+ \hat{b}_n,$$

where $\hat{b}_n$ are the bosonic annihilation operators of the bath. The environment is completely characterised by the spectral function $\mathcal{J}(\omega) = \pi \sum_n \lambda_n^2 \delta(\omega - \omega_n)$ which combines the frequencies of the oscillators $\omega_n$ and their couplings $\lambda_n$. It can be related to the classical reorganization energy associated to the energy relaxation of a sudden electronic transition. This way of modelling a fluctuating dynamical polarization of the environment has found numerous applications to electron transfer problems [9,10].

Solutions of the spin-boson problem can in general display non-Markovian properties [11], dependent on the spectral function. We do not have precise knowledge of $\mathcal{J}(\omega)$ but can reasonably approximate it as having the form $\mathcal{J}(\omega) \propto \omega^s e^{-\omega/\omega_c}$, which for the most relevant electron-phonon interactions to intra- or inter-molecular vibrations is likely to be super-ohmic [12] with $s > 1$ and have a cut-off frequency obeying $U - V > \omega_c > t$. The bath is therefore non-adiabatic with respect to the hopping. Furthermore, the energy gap $U - V \gg t$ means that the model operates in the "large-bias" regime. In this limit the dynamics of the spin-boson model gives the decay rate $\Gamma$ for an initial state $|0\rangle$ to decay to $|g\rangle$ by dissipation into the environment as [8]

$$\Gamma = \frac{1}{2}\left(\frac{t}{U-V}\right)^2 \mathcal{J}(U-V).$$

Only the value of the spectral function at the bias energy $U - V$ to be dissipated is relevant. Since this is above the cut-off the closest high frequency modes are intra-molecular vibrations which are likely to be unaffected by pressure to leading order. We therefore assume that the (unknown) value of $\mathcal{J}(U - V)$ is pressure independent and predict that the recombination rate scales with pressure via the dependent coherent parameters $t$ and $V$ as

$$\frac{\Gamma(P)}{\Gamma(0)} = \left(\frac{t(P)}{t(0)}\right)^2 \left(\frac{U-V(0)}{U-V(P)}\right)^2, \tag{S1}$$

after assuming that $U$ is also independent of pressure. This scaling of the rate sensibly encodes an increase in the decay rate $\Gamma$ with increasing hopping $t$, which is the process ultimately responsible for recombination, and an increase with decreasing bias $U - V$ between the levels.

### E. Master equation for holon-doublon recombination

The super-ohmic and large-bias regime of the spin-boson model is known to exhibit overdamped behaviour [8]. Consequently non-Markovian memory effects, such as recurrences and oscillations in the populations due to back-action of the environment [13], are highly suppressed and its transient behaviour rapidly settles to the Markovian decay. The dynamics of the large-bias super-ohmic regime is therefore effectively Markovian [14] (although not necessarily weak-coupling). Moreover the experimentally measured short-time decay of the reflectivity shown in Fig. 3 of the main text, which is an approximate measure of the carrier density, displays a monotonic profile that is highly indicative of a Markovian process. Based on these observations we account for recombination in the strong-coupling limit effective model by introducing a local Markovian quantum dissipation process which incoherently drives $|0\rangle \to |g\rangle$. This is described by a Lindblad master equation [15] describing the evolution of the density matrix $\rho$ as (taking $\hbar = 1$)

$$\frac{d}{d\tau}\rho(\tau) = -i[H,\rho(\tau)] + \Gamma\left(C\rho(\tau)C^\dagger - \frac{1}{2}C^\dagger C\rho(\tau) - \frac{1}{2}\rho(\tau)C^\dagger C\right),$$

where $C = |g\rangle\langle 0|$ is the jump operator. The resulting dynamics of the initial photo-excited state [16] $\rho(0) = |0\rangle\langle 0|$ then describes the competition between this local dissipation of a bound holon-doublon and the coherent evolution which acts to unbind the pair. We then numerically computed the population $n_g(\tau) = tr(\rho(\tau)|g\rangle\langle g|)$ in the ground state as a function of time up to 4 ps and found it was well fitted with a trial function $n_g(\tau) = A\left(1 - e^{-\Gamma_{eff}\tau}\right)$ allowing a determination of the effective quasi-particle decay rate $\Gamma_{eff} < \Gamma$. To predict the pressure dependence we first fitted the bare rate $\Gamma(0)$ so that $\Gamma_{eff}(0)$ equalled the experimental value for zero pressure. From this we used the spin-boson scaling in Eq. (S1) to determine $\Gamma(P)$ for all other pressure values from which the numerical solution of the model gave $\Gamma_{eff}(P)$. In Fig. 4 of the main text $\Gamma_{eff}(P)$ is shown for evolution computed by using $L = 20$ to model the particle-hole continuum.